\documentclass[10pt,conference]{IEEEtran}
\usepackage{cite}
\usepackage{amsmath,amssymb,amsfonts}
\usepackage{graphicx, balance}
\usepackage{textcomp}
\usepackage{xcolor}
\def\BibTeX{{\rm B\kern-.05em{\sc i\kern-.025em b}\kern-.08em
    T\kern-.1667em\lower.7ex\hbox{E}\kern-.125emX}}
\begin{document}

\title{Harmonic Retrieval of CFO and Frame Misalignment for OFDM-based Inter-Satellite Links}

\author{\IEEEauthorblockN{Rıfat Volkan Şenyuva}
\IEEEauthorblockA{\textit{Maltepe University} \\
\textit{Electrical-Electronics Engineering}\\
Istanbul, Turkey \\
rifatvolkansenyuva@maltepe.edu.tr}
\and
\IEEEauthorblockN{Güneş Karabulut Kurt}
\IEEEauthorblockA{\textit{Poly-Grames Research Center} \\
\textit{Electrical Engineering,  Polytechnique Montr\'eal}\\
Montr\'eal, Canada \\
gunes.kurt@polymtl.ca}
}

\maketitle

\begin{abstract}
As dense low Earth orbit (LEO) constellations are being planned, the need for accurate synchronization schemes in high-speed environments remains a challenging problem to tackle. To further improve synchronization accuracy in channeling environments, which can also be applied in the LEO networks, we present a new method for estimating the carrier frequency offset (CFO) and frame misalignment in orthogonal frequency division multiplexing (OFDM) based inter-satellite links. The proposed method requires the transmission of pilot symbols to exploit 2-D estimation of signal parameters via rotational invariance techniques (ESPRIT) and estimate the CFO and the frame misalignment. The Cramer-Rao lower bounds (CRLB) of the joint estimation of the CFO and frame misalignment are also derived. Numerical results show that the difference between the proposed method and the state-of-art method is less than 5dB at its worse.
\end{abstract}


\section{Introduction}
\label{sec:intro}

The standardization activities in the integrated 5G networks and low earth orbits (LEO) mobile satellite communication (SatCom) systems and the emerging plans about LEO constellations, introduce a renewed research interest in the performance of inter-satellite links (ISLs) \cite{LEO_5G,ultraLEO,LEO_beyond5G}. 
With the technological advances in small satellites, such as cubesats, ISLs will serve as the essential link to enable the satellites to work together to accomplish the same task and satellite formation flying missions \cite{surveyISL,cubesat,interSatSynch}. 

Orthogonal frequency division multiplexing (OFDM) emerges as the leading waveform candidate for ISLs due to the inherent advantages to provide high data rates at relatively low complexity. OFDM is a multicarrier modulation technique widely adopted by modern mobile communication systems such as WiFi networks 
(IEEE 802.11a-802.11be),  
Long Term Evolution (LTE), and LTE-Advanced. It is also considered as the basic waveform for the 5G New Radio (NR). Although OFDM has very desirable traits including high spectral efficiency, achievable high data rates, and robustness in the presence of multipath fading, it is also very sensitive to synchronization errors such as the carrier frequency offset (CFO) and frame misalignments \cite{phyLayerOFDM}. Due to the high speed movement of the LEO satellites, synchronization is expected to be a major challenge in ISLs with the high Doppler shifts \cite{LEO_5G,LEODoppler}.  
In the presence of a high Doppler shift, the subcarriers of an OFDM system with CFO may no longer be orthogonal and the resulting inter-carrier interference (ICI) may degrade the performance of the system. Another synchronization issue arises when the starting position of the discrete Fourier transform (DFT) window at the receiver is misaligned. Depending on the position of the misalignment, the effects of the frame misalignment may range from a simple phase offset to ICI.

Although the research on carrier and frame timing synchronization for OFDM systems is very mature \cite{optimalMaxLikely,timingFreqSynch,jointCarrSamplHarmonic,espritCFO,zadoffChu,LEOSynch}, in this paper, we aim to provide a fresh perspective to OFDM's synchronization problem by also aiming to combat the residual Doppler shifts that may be encountered in ISLs due to their high mobility. Estimating the Doppler shifts is not only important for dealing with CFO effects but the Doppler shifts can also be used for securing the ISLs between spacecrafts, by using them as a mutually shared secret \cite{ozan}.

The proposed method transforms the estimation of the CFO and the frame misalignment in an OFDM based inter-satellite link into a 2-D harmonic retrieval problem. The representation is in the frequency domain and this allows the estimation of the frame misalignment unlike \cite{jointCarrSamplHarmonic,espritCFO} where the frame misalignment is assumed to be corrected prior. When compared to the methods that utilize the cyclic prefix (CP) portion of the OFDM symbol \cite{optimalMaxLikely,LEOSynch}, the proposed method relies on the use of pilot symbols. Furthermore, the proposed method does not require the knowledge of the noise statistics unlike the methods that use the CP  \cite{optimalMaxLikely,LEOSynch}. We also derive the Cram\'er-Rao lower bound (CRLB)   for the joint estimation of the CFO and frame misalignment. Numerical results show that for the error range of $[10^{-2},10^{-4}]$, the difference between the proposed 2-D ESPRIT based method and the PSS method \cite{LEOSynch} is less than 5dB at its worse.

The contributions of the paper can be summarized as follows:
\begin{itemize}
	\item The proposed method represents the estimation of the CFO and the frame misalignment in an OFDM-based ISL as a 2-D harmonic retrieval problem. Unlike \cite{jointCarrSamplHarmonic}, the representation is in the frequency domain and this allows the estimation of the frame misalignment. 
	\item When compared to the methods that utilize the cyclic prefix symbols \cite{optimalMaxLikely,LEOSynch}, the proposed method relies on the use of pilot symbols. So the length of the CP symbols is not a factor that affects the estimation performance of the proposed method.
	\item The proposed method does not require knowledge of noise statistics, unlike the methods that use cyclic prefix symbols
	\item A disadvantage of the proposed method is that due to relying on sending constant pilot symbols, the peak-to-average ratio is high. 
 \cite{optimalMaxLikely,LEOSynch}. 
\end{itemize}

The organization for the rest of this paper is in the following way; Section 2 introduces the signal model for the OFDM-based ISLs. Section 3 reformulates the signal model as a 2-D estimation problem. Section 4 shows the CRLB of the parameters in the reformulated model. We compare the simulation results of the proposed method against the state-of-art methods in Section 5. Finally, Section 6 presents the conclusions and directions for future work.

\section{Signal Model}
\label{sec:signal}

In an OFDM system, the symbols are transmitted with the sampling period $T_{s}$ in a series of frames denoted by $X_{m}[k]$ where $m$ indicates the $m$-th OFDM frame and $k$ is the subcarrier index. The time-domain symbols are modulated by applying the inverse DFT to the frequency-domain symbols for $k=0,\ldots,N-1$ where $N$ is the total number of subcarriers. Then $N_{g}$ number of CP samples are appended to the front of the time-domain frame as in
\begin{equation}
    x_{m}[n]=\frac{1}{N}\sum_{k=0}^{N-1}X_{m}[k]e^{i2\pi k(n-N_{g})/N}
    \label{eq:OFDMblock}
\end{equation}
where $0\le n \le N_{t}-1$ and $N_{t}=N+N_{g}$. The discrete-time received frames in baseband can be written as 
\begin{equation}
	r_{m}[n]=e^{i2\pi \epsilon mN_{t}/N}e^{i2\pi \epsilon n/N}(h_{m}\circledast\tau_{p}x_{m})[n]+z_{m}[n]
	\label{eq:ISI}
\end{equation}
where $\circledast$ denotes the cyclic convolution operation, $h_{m}[n]$ is the overall channel that is the result of the convolution of the multipath fading channel and the pulse shaping filters of both the transmitter and the receiver, the starting position of the DFT window at the receiver is shown as $\tau_{p}x_{m}[n]=x_{m}[n+p]$, and the additive white Gaussian noise (AWGN), $z_{m}[n]$, is modeled as a circularly symmetric Gaussian random variable, i.e. $z_{m}[n]\sim\mathcal{CN}(0,N_{0})$. Due to the clock inaccuracies of the transmitter and the receiver oscillators, and the Doppler spread caused by the mobility of the satellites, the conversion from passband to baseband generates the unwanted multiplicative terms, $e^{i2\pi \epsilon mN_{t}/N}e^{i2\pi \epsilon n/N}$, in   \eqref{eq:ISI} where $\epsilon$ is the total CFO term normalized by the subcarrier spacing, $1/NT_{s}$, as
\begin{equation}
    \epsilon=NT_{s}(f_{d}-\Delta_{f_{c}}).
    \label{eq:totalCFO}
\end{equation}
$\Delta_{f_{c}}$ and $f_{d}$ in \eqref{eq:totalCFO} shows the clock difference and the Doppler spread, respectively. Since the free-space loss and thermal noise of the electronics are enough to characterize the ISLs, an AWGN channel model, i.e. $h_{m}[n]=\delta_{0}[n]$, can be used in \eqref{eq:ISI} \cite{cubesat}. Thus the received signal for the discrete baseband equivalent model can be written as
\begin{equation}
    r_{m}[n]= e^{i2\pi \epsilon m(1+\alpha)}e^{i2\pi \epsilon n/N}x_{m}[n+p]+z_{m}[n]
    \label{eq:AWGN}
\end{equation}
where $\alpha=N_{g}/N$. The normalized CFO, $\epsilon=\varepsilon+\ell$, consists of a fractional part, i.e. $|\varepsilon|\le 0.5$, and an integer part $\ell\in\mathbb{Z}^{+}$. The OFDM receiver first removes the cyclic prefix samples and then applies the DFT to the remaining samples resulting in
\begin{equation}
    R_{m}[k]=C_{m}[k]\circledast e^{i2\pi(\ell-k)p/N}X_{m}[k-\ell]+Z_{m}[k],
    \label{eq:DFTapplied} 
\end{equation}
where $C_{m}[k]$ is given as
\begin{IEEEeqnarray}{rCl}
	C_{m}[k]&=&\frac{\sin(\pi[\varepsilon-k])}{N\sin(\pi[\varepsilon-k]/N)}e^{i\pi[\varepsilon-k](N-1)/N} \nonumber \\
	              &\times& e^{i2\pi m[\varepsilon(1+\alpha)+\ell\alpha]}e^{i2\pi (\varepsilon+\ell) \alpha}
\end{IEEEeqnarray}
and $Z_{m}[k]$ is also circularly symmetric Gaussian that is $Z_{m}[k]\sim\mathcal{CN}(0,NN_{0})$ since the DFT is a linear transformation and the circular symmetry is invariant to linear transformations.

\section{Estimation of CFO and Frame Misalignment Using 2-D ESPRIT}
\label{sec:2D_ESPRIT}

The DFT of the received samples, $R_{m}[k]$, in \eqref{eq:DFTapplied} can be rewritten as a 2-D signal model by using the OFDM frame index, $m$, as a second dimension taking values in the range $0\le m\le M-1$
\begin{IEEEeqnarray}{rCl}
	R[m,k]&=& e^{i2\pi f_{1}m}e^{i2\pi f_{2}k}e^{i\psi}\sum_{r=0}^{N-1}\frac{\sin(\pi(\varepsilon-r))}{N\sin(\pi[\varepsilon-r]/N)} \nonumber \\
	         && e^{i2\pi r \left( \frac{1-N}{2N}+\frac{p}{N} \right)}X[m,k-\ell-r]+Z[m,k], \label{eq:DFTappliedRewrite1}
\end{IEEEeqnarray}
where the frequencies and the phase terms are respectively
\begin{IEEEeqnarray}{rCl}
	f_{1}&=&\varepsilon(1+\alpha)+\ell\alpha \label{eq:freq1} \\
	f_{2}&=&-p/N  \label{eq:freq2} \\
	\psi&=&2\pi\left\lbrack (\varepsilon+\ell) \alpha+\frac{\ell p}{N}+\varepsilon\frac{N-1}{2N} \label{eq:phase} \right\rbrack.
\end{IEEEeqnarray}
The transmitted OFDM symbols in \eqref{eq:DFTappliedRewrite1}, $X[m,k-\ell-r]$, depend on both the frame index, $m$, and the subcarrier index, $k$. This dependency can be removed by sending the same pilot symbol on each subcarrier that is $X[m,k]=X$ for all $m$ and $k$, and then the DFT of the received samples \eqref{eq:DFTappliedRewrite1} can be formulated as a harmonic retrieval problem
\begin{equation}
    R[m,k]=c \, \phi^{m}\theta^{k}+Z[m,k] , \label{eq:2DParameterEstimation}
\end{equation}
where $\phi=e^{i2\pi f_{1}}$, $\theta=e^{i2\pi f_{2}}$ and $c$ is a complex coefficient
\begin{equation}
	c=e^{i\psi}X\sum_{r=0}^{N-1}\frac{\sin(\pi(\varepsilon-r))}{N\sin(\pi[\varepsilon-r]/N)}e^{i2\pi r \left( \frac{1-N}{2N}+\frac{p}{N} \right)}.
\end{equation}
The signal in \eqref{eq:2DParameterEstimation} has a single 2-D mode defined by the frequencies $\lbrace f_{1},f_{2} \rbrace$, and a complex coefficient $c=\lambda e^{i\varphi}$ where $\lambda=|c|$ and $\varphi=\angle c$. The ESPRIT-based estimation methods identify the pairs $\lbrace f_{1},f_{2} \rbrace$ from the observed data $R[m,k]$ by turning the 2-D estimation problem into two 1-D estimation problems and exploiting the shift-invariance structure of the signal subspace for each 1-D problem \cite{2D_ESPRIT}. Since the unknown 2-D signal mode is undamped, the forward-backward prediction \cite{2D_ESPRIT} can be applied to increase the estimation accuracy by using an extended data matrix $\mathbf{R}_{ee}$ as
\begin{equation}
    \mathbf{R}_{ee}=[\mathbf{R}_{e} \quad \mathbf{\Pi}\,\mathbf{R}^{*}_{e}\,\mathbf{\Pi}],
    \label{eq:extendedDataMatrix}
\end{equation}
where complex conjugation without transposition is shown by the $(\cdot)^{*}$ symbol, and $\mathbf{\Pi}$ is a permutation matrix with ones on its antidiagonal and zeroes elsewhere. $\mathbf{R}_{e}$ \eqref{eq:extendedDataMatrix} is the enhanced Hankel block structured matrix that is constructed by applying an observation window of size $P\times Q$ through the rows of the noisy received samples of which one of the dimensions is fixed. $\mathbf{R}_{e}$ matrix for the first dimension is given as
\begin{equation}
    \mathbf{R}_{e1}=\left\lbrack
                    \begin{array}{cccc}
                         \mathbf{R}_{(0)} & \mathbf{R}_{(1)} & \cdots & \mathbf{R}_{(M-P)} \\
                         \mathbf{R}_{(1)} & \mathbf{R}_{(2)} & \cdots & \mathbf{R}_{(M-P+1)} \\
                         \vdots & \ddots & \ddots & \vdots \\
                         \mathbf{R}_{(P-1)} & \mathbf{R}_{(P)} & \cdots & \mathbf{R}_{(M-1)}
                    \end{array}
                   \right\rbrack,
\end{equation}
where $\mathbf{R}_{(m)}$ is a Hankel matrix of size $Q\times (N-Q+1)$ as given below
\begin{equation}
    \mathbf{R}_{(m)}=\left\lbrack
                        \begin{array}{ccc}
                             R[m,0] & \cdots & R[m,N-Q] \\
                             R[m,1] & \cdots & R[m,N-Q+1] \\
                             \vdots & \ddots & \vdots \\
                             R[m,Q-1] & \cdots & R[m,N-1]
                        \end{array}
                     \right\rbrack.
\end{equation}
The extended data matrix constructed according to \eqref{eq:extendedDataMatrix} for the first dimension, $\mathbf{R}_{ee1}$, can be decomposed in terms of a signal and a noise subspace as
\begin{equation}
    \mathbf{R}_{ee1}=c\,\mathbf{s}_{L1}\mathbf{s}^{T}_{R1}+\mathbf{Z}_{1},
    \label{eq:extendedData}
\end{equation}
where $\mathbf{Z}_{1}$ is the Hankel block structured matrix constructed from the noise samples $Z[m,k]$ in the same way as $\mathbf{R}_{e1}$ from $R[m,k]$. $\mathbf{s}_{L1}$ is a vector of size $PQ\times 1$ given as $\mathbf{s}_{L1}=[\boldsymbol{\theta}_{1} \, \boldsymbol{\theta}_{1}\phi \, \ldots \boldsymbol{\theta}_{1}\phi^{P-1}]^{T}$ where $\boldsymbol{\theta}_{1}$ is $\boldsymbol{\theta}_{1}=\lbrack 1 \,\,\theta \,\ldots\,\theta^{Q-1} \rbrack^{T}$. The construction of the vector $\mathbf{s}_{R1}$ is similar to that of $\mathbf{s}_{L1}$. The rank of the signal subspace, $c\,\mathbf{s}_{L1}\mathbf{s}^{T}_{R1}$, must be equal to the number of signal components which is equal to one since there is only one user. In order for the rank of the signal part be at least equal to one, $P$ and $Q$ must satisfy the following inequalities \cite{2D_ESPRIT}
\begin{equation}
    M \ge P \ge 1, \qquad N \ge Q \ge 1.
    \label{eq:obsWindSize}
\end{equation}
The singular value decomposition (SVD) of $\mathbf{R}_{ee1}$ yields
\begin{equation}
    \mathbf{R}_{ee1}=\mathbf{U}_{S1}\mathbf{\Sigma}_{S1}\mathbf{V}^{H}_{S1}+\mathbf{U}_{Z1}\mathbf{\Sigma}_{Z1}\mathbf{V}^{H}_{Z1}.
    \label{eq:SVD}
\end{equation}
While the singular vectors and the singular value of the signal mode is contained respectively in $\mathbf{U}_{S1}$, $\mathbf{V}^{H}_{S1}$, and $\mathbf{\Sigma}_{S1}$, the singular vectors and the singular values of the remaining noise components are respectively in $\mathbf{U}_{Z1}$, $\mathbf{V}^{H}_{Z1}$, and $\mathbf{\Sigma}_{Z1}$ \eqref{eq:SVD}. 

The frequencies related to the first dimension, $f_{1}$, can be estimated the $\mathbf{F}_{1}$ matrix
\begin{equation}
    \mathbf{F}_{1}=\underline{\mathbf{U}}^{\dagger}_{S1}\overline{\mathbf{U}}_{S1}
    \label{eq:eigs1stDim}
\end{equation}
where $\underline{\mathbf{U}}_{S1}$ (resp. $\overline{\mathbf{U}}_{S1}$) is constructed with the first (resp. last) $(P-1)Q$ rows of the matrix $\mathbf{U}_{S1}$. $\underline{\mathbf{U}}^{\dagger}_{S1}$ denotes the pseudo-inverse matrix of $\underline{\mathbf{U}}_{S1}$ and the total least squares can be applied to calculate $\mathbf{F}_{1}$ in \eqref{eq:eigs1stDim}. The frequency $f_{1}$ is the eigenvalue of the matrix $\mathbf{F}_{1}$ that is
\begin{equation}
    f_{1}=\mathbf{T}_{1}\mathbf{F}_{1}\mathbf{T}^{-1}_{1},
\end{equation}
where $\mathbf{T}_{1}$ is the eigenvector matrix that diagonalizes $\mathbf{F}_{1}$. For the frequency related to the second dimension, $f_{2}$, the $\mathbf{F}_{2}$ matrix can be obtained by using the extended data matrix for the second dimension, $\mathbf{R}_{ee2}$, and following the same steps. Finally the pairing method of \cite{2D_ESPRIT} must also be employed to calculate the eigenvalue decomposition of a linear combination of $\mathbf{F}_{1}$ and $\mathbf{F}_{2}$
\begin{equation}
    \beta \mathbf{F}_{1}+(1-\beta)\mathbf{F}_{2}=\mathbf{T}\mathbf{\Sigma}\mathbf{T}^{-1},
\end{equation}
where $\beta$ is a scalar. The diagonalizing transformation $\mathbf{T}$ is applied to both $\mathbf{F}_{1}$ and $\mathbf{F}_{2}$
\begin{IEEEeqnarray}{rCl}
    f_{1}&=&\mathbf{T}\mathbf{F}_{1}\mathbf{T}^{-1} \\
    f_{2}&=&\mathbf{T}\mathbf{F}_{2}\mathbf{T}^{-1}
\end{IEEEeqnarray}
which yields the ordered frequencies.

\section{Cram\'er-Rao lower bound (CRLB) Analysis}
\label{sec:CRLB}

The unknown parameters of \eqref{eq:2DParameterEstimation} can be collected in a vector of size $4\times 1$ as
\begin{equation}
    \boldsymbol{\vartheta}=\left\lbrack \omega_{1} \, \omega_{2} \,\, \lambda \,\, \varphi \right\rbrack^{T}.
\end{equation}
where $\omega_{1}=2\pi f_{1}$ and $\omega_{2}=2\pi f_{2}$ represent the angular frequencies. If the DFT of the received signal samples \eqref{eq:2DParameterEstimation} are written as an $MN\times 1$ column vector
\begin{eqnarray}
    \mathbf{r}=\lbrack R[0,0],\ldots,R[0,N-1],\ldots \nonumber \\
                       R[M-1,0],\ldots R[M-1,N-1] \rbrack^{T}
\end{eqnarray}
then the joint probability density function (PDF) of the multivariate circularly symmetric Gaussian random vector $\mathbf{r}\sim\mathcal{CN}(\boldsymbol{\mu},NN_{0}\mathbf{I})$ is given as
\begin{equation}
    f(\mathbf{r})=\frac{1}{(NN_{0}\pi)^{MN}}\exp\left\lbrace -\frac{1}{NN_{0}}\left(\mathbf{r}-\boldsymbol{\mu}\right)^{H}\left(\mathbf{r}-\boldsymbol{\mu}\right) \right\rbrace
    \label{eq:jointPDF}
\end{equation}
where $\mathbf{I}$ is an identity matrix of size $MN\times MN$. The $j$-th entry of the mean vector $\boldsymbol{\mu}$ \eqref{eq:jointPDF} is
\begin{equation}
    \boldsymbol{\mu}_{j}=c\,\phi^{m'_{j}}\,\theta^{k'_{j}}, \qquad j=1,\ldots,MN,
\end{equation}
where
\begin{IEEEeqnarray}{rCl}
    m'_{j}&=&\left\lfloor{(j-1)/N}\right\rfloor\mod M \\
    k'_{j}&=&\left\lfloor j-1 \right\rfloor\mod N.
\end{IEEEeqnarray}
The $(j,j')$ entry of the Fisher information matrix for the multivariate Gaussian PDF \eqref{eq:jointPDF} is 
\begin{equation}
    \mathbf{W}_{j,j'}=\frac{2}{NN_{0}}\Re\left\lbrace\left\lbrack\frac{\partial\boldsymbol{\mu}}{\partial\boldsymbol{\vartheta}_{j}} \right\rbrack^{H} \left\lbrack \frac{\partial\boldsymbol{\mu}}{\partial\boldsymbol{\vartheta}_{j'}} \right\rbrack \right\rbrace,
\end{equation}
where $\Re\lbrace \cdot \rbrace$ takes the real parts of the entries of the matrix \cite{SSA_Multidim}. The diagonal entries of the inverse of the Fisher matrix are the CRLB of the parameters. The CRLB for $\omega_{1}$ and $\omega_{2}$ are written respectively in \eqref{eq:CRBfreq1} and \eqref{eq:CRBfreq2}.
\begin{IEEEeqnarray}{rCl}
    \text{CRLB}(\omega_{1})&=& \frac{6NN_{0}}{c^{2}MN(M^2-1)} 
    \label{eq:CRBfreq1} \\
    \text{CRLB}(\omega_{2})&=& \frac{6NN_{0}}{c^{2}MN(N^{2}-1)} 
    \label{eq:CRBfreq2}
\end{IEEEeqnarray}

\section{Numerical Results}
\label{sec:numRes}

The simulated inter-satellite communication system uses the OFDM modulation which consists $N=64$ subcarriers and $N_{g}=16$ CP samples. The CFO is generated independently for each OFDM frame as a uniform random variable $\epsilon \sim \mathcal{U}[0.2,0.25]$ and the frame misalignment is fixed at $p=2$. The signal-to-noise (SNR) ratio is defined as $\text{SNR}=E_{s}/N_{0}$ where the symbol energy is normalized to unity, $E_{s}=1$. The performance of the proposed method is measured by calculating the mean squared error (MSE) for both the CFO and frame misalignment estimations and a total of 2000 Monte Carlo simulations are run for each SNR value.

\begin{figure}[tb]
    \begin{minipage}[b]{1.0\linewidth}
        \centering
        \centerline{\includegraphics[width=8.8cm]{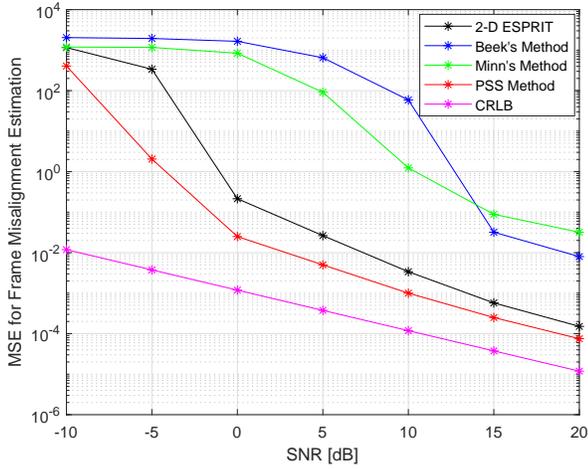}}
    \end{minipage}
    \caption{Comparison of the frame misalignment estimation performance of the proposed 2-D ESPRIT method with Beek's, Minn's, and the PSS methods in the SNR range from -10 dB to 20 dB.}
    \label{fig:MSE_STO}
\end{figure}

The selected methods are the maximum likelihood estimator, also known as Beek's method \cite{optimalMaxLikely} that allows the transmission of data symbols by using the CP samples for estimation, Minn's improved Schmidl \& Cox estimator \cite{timingFreqSynch}, and the cross-correlation based method that utilizes primary synchronization symbols (PSS method) \cite{LEOSynch}. The parameters for the proposed method are as follows; the number of the pilot symbols is $M=2$, the size of the observation window applied to the received samples is chosen according to \eqref{eq:obsWindSize} as $P=Q=2$ and the pairing coefficient is $\beta=8$. The results of the frame misalignment estimation (Figure \ref{fig:MSE_STO}) show that while Beek's method and Minn's method perform worse than the proposed method at each SNR value, the performance of the PSS method surpasses the performance of the proposed method.

\begin{figure}[tb]
    \begin{minipage}[b]{1.0\linewidth}
        \centering
        \centerline{\includegraphics[width=8.8cm]{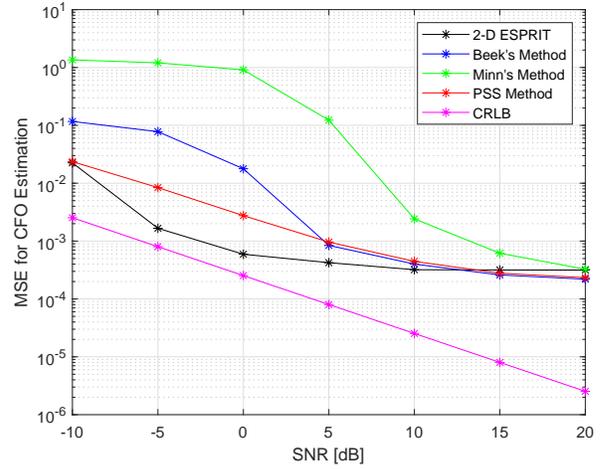}}
    \end{minipage}
    \caption{Comparison of the CFO estimation performance of the proposed 2-D ESPRIT method with Beek's, Minn's, and the PSS methods in the SNR range from -10 dB to 20 dB.}
    \label{fig:MSE_CFO}
\end{figure}

The MSE regarding the CFO estimation results is shown in Figure \ref{fig:MSE_CFO}. The MSE of the proposed 2-D ESPRIT method is closest to the CRLB bound at 0 dB SNR and for the SNR values less than 10 dB, the proposed method performs superior compared to the rest of the methods. The PSS method is the second-best performer for SNR values less than 5 dB. 

The computational complexities of all the methods and their numerical evaluations for the parameter values used in the simulations are in Table \ref{tab:complexityComparison}. The complexities are in terms of real floating-point operations (flops). One complex multiplication is counted as 6 real flops while one complex addition is counted as 2 real flops. The SVD step determines the complexity of the proposed 2-D ESPRIT estimation method and the complexity is written in terms of the size of the extended data matrix, $\mathbf{R}_{ee}\in\mathbb{C}^{L\times K}$, where $L=PQ=4$ and $K=2(N-Q+1)(M-P+1)=126$. The symbol $G$ used in the complexity of the PSS method shows the size of the search grid that is used in the CFO estimation step. $D$, the number of repetition of the training symbol used in Minn's method, is $D=4$. In the numerical results, $M$ and $G$ are chosen as $M=2$ and $G=500$ respectively. While the complexity of the proposed method is less than the PSS's method, the complexities of both Beek's and Minn's methods are lower.

\begin{table}[tbp]
\renewcommand{\arraystretch}{1.3}
\caption{Complexity Comparison of the Methods}
\begin{center}
\begin{tabular}{|c|c|c|}
\hline
\textbf{Method}& \textbf{Complexity} & \textbf{Number} \\
\hline
Proposed Harmonic Retrieval & $2LK^{2}+K^{3}+K+LK$ & 2127384 \\
\hline
PSS \cite{LEOSynch} & $(NK+G)(15N+5)$ & 729540 \\
\hline
Beek \cite{optimalMaxLikely} & $24N_{t}N_{g}+10N_{t}$ & 31520 \\
\hline
Minn's \cite{timingFreqSynch} & $36(N^{2}/D)+6N$ & 36480 \\
\hline
\end{tabular}
\label{tab:complexityComparison}
\end{center}
\end{table}


\section{Conclusion}
\label{sec:con}
Synchronization in the ISLs of the LEO systems is a critical issue due to the Doppler spread caused by the high mobility of the satellites. We presented a novel method for the estimation of the CFO and frame misalignment in OFDM-based ISLs. We reformulate the synchronization problem as a 2-D harmonic retrieval problem in the frequency domain and apply the 2-D ESPRIT method to estimate the parameters. Since this new approach is in the frequency domain unlike the previous harmonic retrieval approach, this allows the joint estimation of the CFO and the frame misalignment.

When compared to other synchronization methods that also rely on the pilot symbols like the well-known Schmidl \& Cox method and the PSS method, the proposed method requires transmitting frames of pilot symbols for the reformulation to work. The CRLB for the joint estimation of the CFO and the frame misalignment is derived and the performance of the proposed method is compared against the Beek's, Schmidl \& Cox, and the PSS methods. Numerical results show that respectable estimation performance can be achieved by using the proposed method with only two consecutive pilot frames.
\balance
\bibliographystyle{IEEEbib}
\bibliography{IEEEabrv,refs}

\end{document}